\documentclass[structabstract]{aa}
\usepackage{graphicx,rotate,psfig,subfigure,supertabular,times}
\usepackage{txfonts}
\begin{document}
   \title{Understanding BL Lac objects}
  \subtitle{Structural \& kinematic mode changes in the BL Lac object PKS 0735+178}

   \author{S. Britzen
          \inst{1}
      \and
      A. Witzel \inst{1}
          \and
      B.P. Gong
          \inst{2}
      \and J.W. Zhang \inst{2}
      \and Gopal-Krishna \inst{3}
      \and Arti Goyal \inst{4}
      \and M.F. Aller\inst{5}, H.D. Aller\inst{5}
      \and J.A. Zensus \inst{1}
          }

   \institute{Max-Planck-Institut f\"ur Radioastronomie, Auf dem H\"ugel 69,
              D-53121 Bonn\\
              \email{sbritzen@mpifr-bonn.mpg.de}
          \and Department of Physics, Huazhong University of
          Science and Technology, Wuhan 430074, P.R. China
          \and National Centre for Radio Astrophysics/TIFR, Post Bag 3, Pune University Campus, Pune, 411007, India
	  \and Aryabhatta Research Institute of observational sciencES (ARIES), Manora Peak, Naini Tal 263 129, India
          \and Astronomy Department, University of Michigan, Ann Arbor, MI 48109-1042, USA
           \\
             }

   \date{Received \today; accepted \today}
  \abstract
   {We present evidence that parsec-scale jets in BL Lac objects may be
significantly distinct in kinematics from their counterparts in quasars. 
}
   {A detailed study of the jet components' 
motion reveals that the standard AGN paradigm of apparent superluminal 
motion does not always describe the kinematics in BL Lac objects. We study 
0735+178 here to augment and improve the understanding of the peculiar 
motions in the jets of BL Lac objects as a class.}
   {We analyzed 15 GHz VLBA (Very Long Baseline Array) observations (2cm/MOJAVE survey) performed 
at 23 epochs between 1995.27 and 2008.91. We then compared the jet kinematics with the optical and radio light curves
available for this BL Lac object and point out some striking correlations 
between the properties of the radio knots and the features in the light 
curves. }
    {We found a drastic structural mode change in the VLBI jet of 0735+178,
between 2000.4 and 2001.8 when its twice sharply bent trajectory turned 
into a linear shape. We further found that this jet had undergone a similar 
transition sometime between December 1981 and June 1983. A mode change, 
occurring in the reverse direction (between mid-1992 and mid-1995) has already 
been reported in the literature.
These structural mode changes are found to be reflected in changed
kinematical behavior of the nuclear jet, manifested as an apparent 
superluminal motion and stationarity of the radio knots. 
In addition, we found the individual mode changes to correlate in time 
with the maxima in the optical light curve. 
}
   {We found for this BL Lac 
object a drastic change in the kinematics of the nuclear jet, i.e, transition 
from 'typical superluminal' to an unusual 'stationary' state. Interestingly, 
we found that this change is accompanied with a mode change in the nuclear
jet's morphology. The long sequences of VLBA images reported here for
this BL Lac object do not support the commonly assumed connection 
between radio flux-density outbursts and the ejection of new VLBI components. 
}
\maketitle
\section{Introduction}
The strong radio source PKS 0735+178 is an optically bright classical BL Lac 
object, highly variable at radio/optical wavelengths (e.g., Goyal et al. 
2009 and references therein). Based on an absorption feature in the optical
spectrum, a lower redshift limit of 0.424 was established (Abraham, et al. 1991; 
Stickel et al. 1993; Scarpa et al. 2000). 
This is consistent with the non detection of the optical host in the HST 
snapshot imaging survey by Sbarufatti et al.  (2005). A companion (possible 
absorber) has been detected at a projected separation of 22 kpc (Pursimo et 
al. 1999). Extensive monitoring of this blazar in the radio and optical 
bands revealed variability on different timescales. Periodic features of 
13.8-14.2 years were found by applying the Jurkevich method to about a 
century long optical monitoring data (Fan et al. 1997; Qian \& Tao 2004). 
Ciprini et al. (2007) reported possible detection of several shorter 
time scales. In their search for intra-night optical variability on 17 nights 
between 1998 and 2008, Goyal et al. (2009) did not detect on any occasion intra-night variability with an amplitude $\geq$3\%, though such 
amplitudes are typical for radio selected BL Lacs (Gopal-Krishna et al. 2003).  
Gupta et al. (2008) claimed detection of optical variability with an 
amplitude of 4.9\% on one night, but since that light curve remains steady, 
any such variability must be of the flicker type and not on hour-like time scale. \\
In the radio band, an unusual property highlighted for this BL Lac is its 
remarkably flat radio spectrum, termed as ``cosmic conspiracy" (Cotton et al. 
1980). 

\subsection{Radio morphology: kpc- to pc-scale}
PKS 0735+178 has been observed with radio interferometers at various
frequencies. The 1.64 GHz VLA (Very Large Array) observations by Murphy et al. (1993) revealed
a rather compact structure, slightly elongated at a position angle of
$\sim$160 deg. The 5 GHz MERLIN (Multi-Element Radio Linked Interferometer Network) image taken in 1983 is dominated by a 
core and also shows a very weak component to the southeast (B$\aa\aa$th 
\& Zhang 1991). The core-to-component flux density ratio is 200:1 and the 
position angle of the component is $\simeq$160 deg - in agreement with the 
VLA observations by Murphy et al. (1993). These observations established
a large misalignment between the kpc-scale and pc-scale structures of 
this source by $\sim$ 90-100 deg. \\
PKS 0735+178 has been a favorite target of VLBI monitoring programs,
and striking changes in its nuclear jet are revealed from the more 
than two decades of its VLBI monitoring.\\
B$\aa\aa$th \& Zhang (1991) imaged the source at 5 GHz with global VLBI 
arrays at four epochs between 1979 and 1984. The large structural changes 
thus found were interpreted by them in terms of relativistic $(\gamma \sim$ 8) 
motion of the radio knots along a curly path with a steadily decreasing 
viewing angle. Particularly instructive are their VLBI images of the inner 
5 mas region, taken at the epochs 1981.9 and 1983.5, which (despite
modest resolution) reveal a clear changeover from a kinky {\it staircase} 
profile in 1981.9 to a {\it straight} mode in 1983.5 (see below). This 
morphological transition, first noted in the present work, is discussed 
in Sect. \ref{results}.\\ 
Gabuzda et al. (1994) observed the source with the VLBA at the epochs 1987.41 (5 GHz), 1990.47 (8.4 GHz), and 1992.44 (5 GHz). In all these 
observations, the source revealed a straight jet extending to the northeast. 
G$\acute{\rm o}$mez et al. (2001) present VLBA observations at 8.4 and 43 
GHz between August 1996 and May 1998. Between March 1996 and May 2000 Agudo 
et al. (2006) observed the source using the VLBA at frequencies between 5 
and 43 GHz. The observations by G$\acute{\rm o}$mez et al. and Agudo et al. 
show a twisted jet with two sharp apparent bends of 90$\deg$ within two 
milli-arcseconds of the core. Below we call this trajectory {\it staircase} mode. Between mid-1992 and mid 1995, the jet trajectory 
changed again drastically from {\it straight} to {\it staircase} 
(G$\acute{\rm o}$mez et al. 2001).
Unfortunately, no VLBI observations documenting this second transition 
are available.

From early VLBI observations of this BL Lac object B$\aa\aa$th \& Zhang 
(1991) estimated bulk Lorentz factors of up to $\sim 8 h^{-1} c$ for
the knots in the VLBI jet, moving from the core towards a stationary 
component located $\sim 4$ millisecond northeast of the core. A subsequent 
paper by Gabuzda et al. (1994) reports an apparent superluminal motion of 
$\geq$7.4 $h^{-1} c$ of one knot. After mid-1995 the components show a tendency 
to cluster near several jet positions, remaining stationary in time (Sects. 
\ref{results} and \ref{kinematics}).

\subsection{The present work}
The extensive VLBA coverage of this blazar at 15 GHz during 1995-2009 
under the MOJAVE program allowed us to undertake a detailed probe 
of structural transitions in the nuclear jet of this BL Lac object
(Sect. \ref{results}). Thus, in addition to the last documented mode change in 
the pc-scale trajectory (G$\acute{\rm o}$mez et al. 2001), we found 
clear evidence for another mode change between 2000.4 and 2001.8. 
We find that from the {\it staircase} mode observed since mid 1995 the
jet trajectory has reverted to a straight morphology during the time
interval covered by our VLBA data analysis, so that the
timing of this morphological change can be pinned down to an uncertainty 
of merely $\sim$ 1.6 years. In comparison, the uncertainty for the mode change 
inferred from the data of G$\acute{\rm o}$mez et al. was $\sim$ 3 years.
According to them, the morphological transitions are most likely caused by 
pressure gradients in the external medium through which the jet 
propagates. We discuss this hypothesis with regard to the mode transitions
found by us. We further examine these mode transitions gleaned from the 
VLBI data, as mentioned above, with regard to the observed trends in the 
radio and optical light-curves of this intensively monitored blazar. \\

This paper is organized in the following way: we first present the results 
of our analysis of the archival VLBA data (MOJAVE) and point out the newly 
noticed correlations between the morphological and flux-density variations
of this source. We then discuss our findings in the context of previously
reported information about this blazar. 
In particular,  we explore the connection
between the peculiar mode changes and kinematics of its nuclear jet.
Apparent transverse velocities ($\beta_{\rm app}$) in this paper were 
calculated with the one-year WMAP (Wilkinson Microwave Anisotropy Probe) data (Spergel et al., 2003); the values 
adopted are $h=0.71^{+0.04}_{-0.03}$; $\Omega_m h^2=0.135^{+0.008}_{-0.009}$; 
$\Omega_{\rm tot}=1.02\pm0.2$. 
The one-year and three-year WMAP parameters are less than $0.1c$ for $\mu=0.1$ 
mas\,yr$^{-1}$ out to $z=4$; this is small with respect to the formal 
uncertainties in the apparent velocities.

\section{Data analysis}
We analyzed the VLBA observations for 23 epochs, taken at 15 GHz under 
the 2cm/MOJAVE survey between 1995 and 2009.
The data were model-fitted with the {\it difmap} model-fitting routine. 
To each epoch's VLBA image we attempted to fit the minimum number of
components needed to describe the structure adequately to facilitate
a least ambiguous component identification from epoch to epoch.
For each image the model-fitting process started with a point-like source 
model. In order to keep the comparison of the images for different epochs
simple and credible, we assumed the components to be circular; highly elliptical components are
not likely anyway.
\begin{figure*}[htb]
 \centering
 \subfigure[]{\includegraphics[clip,angle=-90,width=8.5cm]{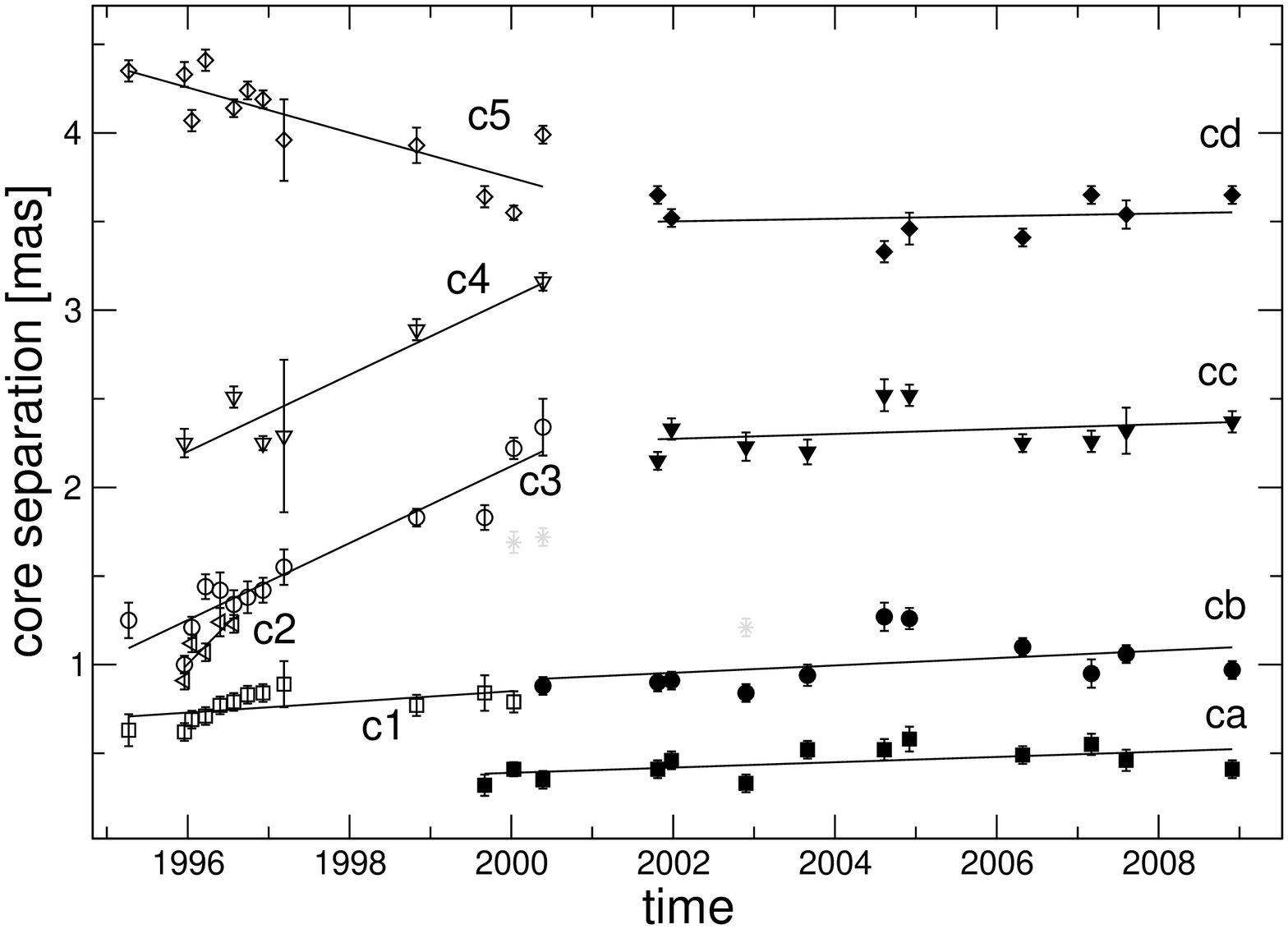}}
   \subfigure[]{\includegraphics[clip,angle=-90,width=8.5cm]{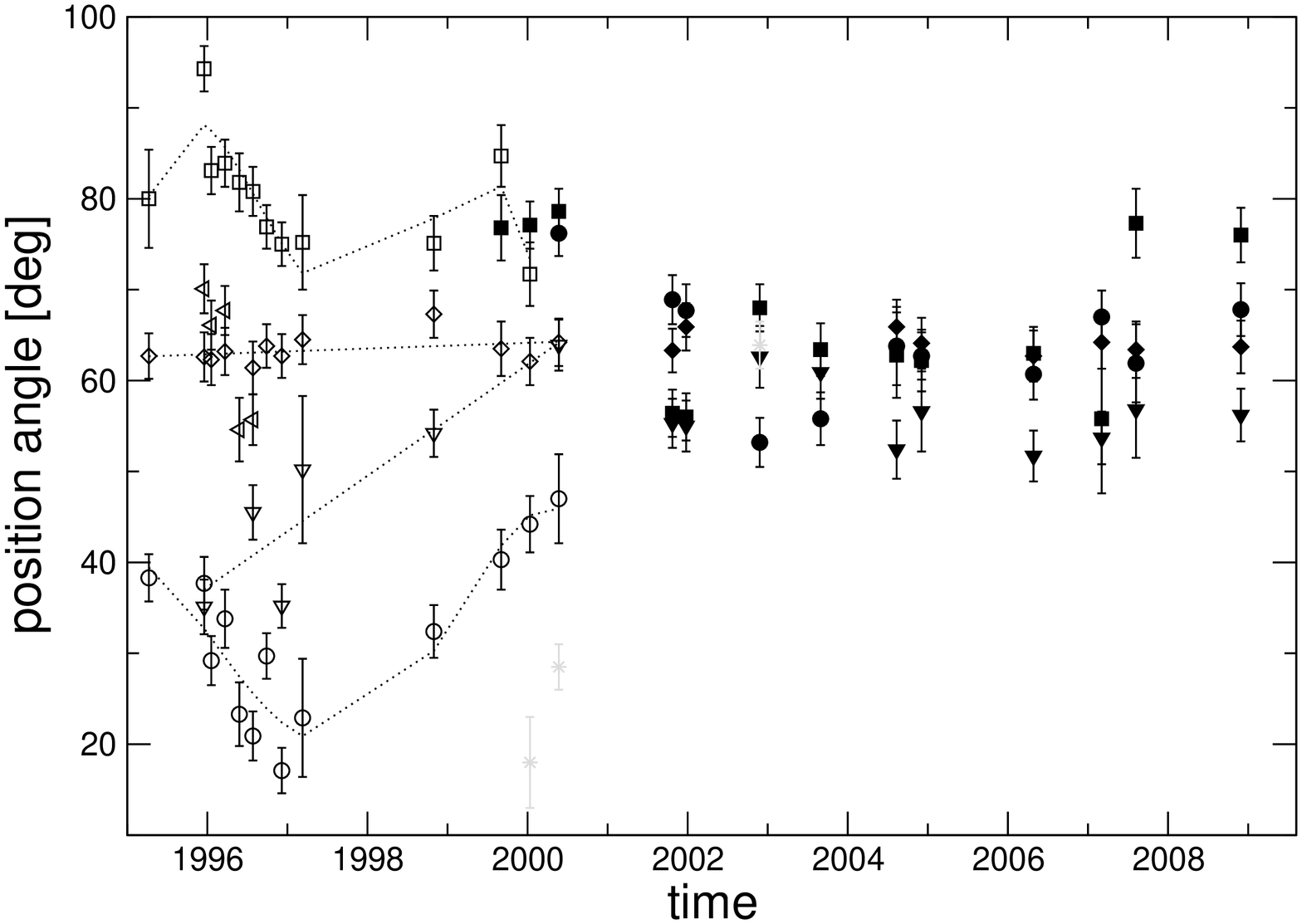}}\\
   \subfigure[]{\includegraphics[clip,angle=-90,width=8.5cm]{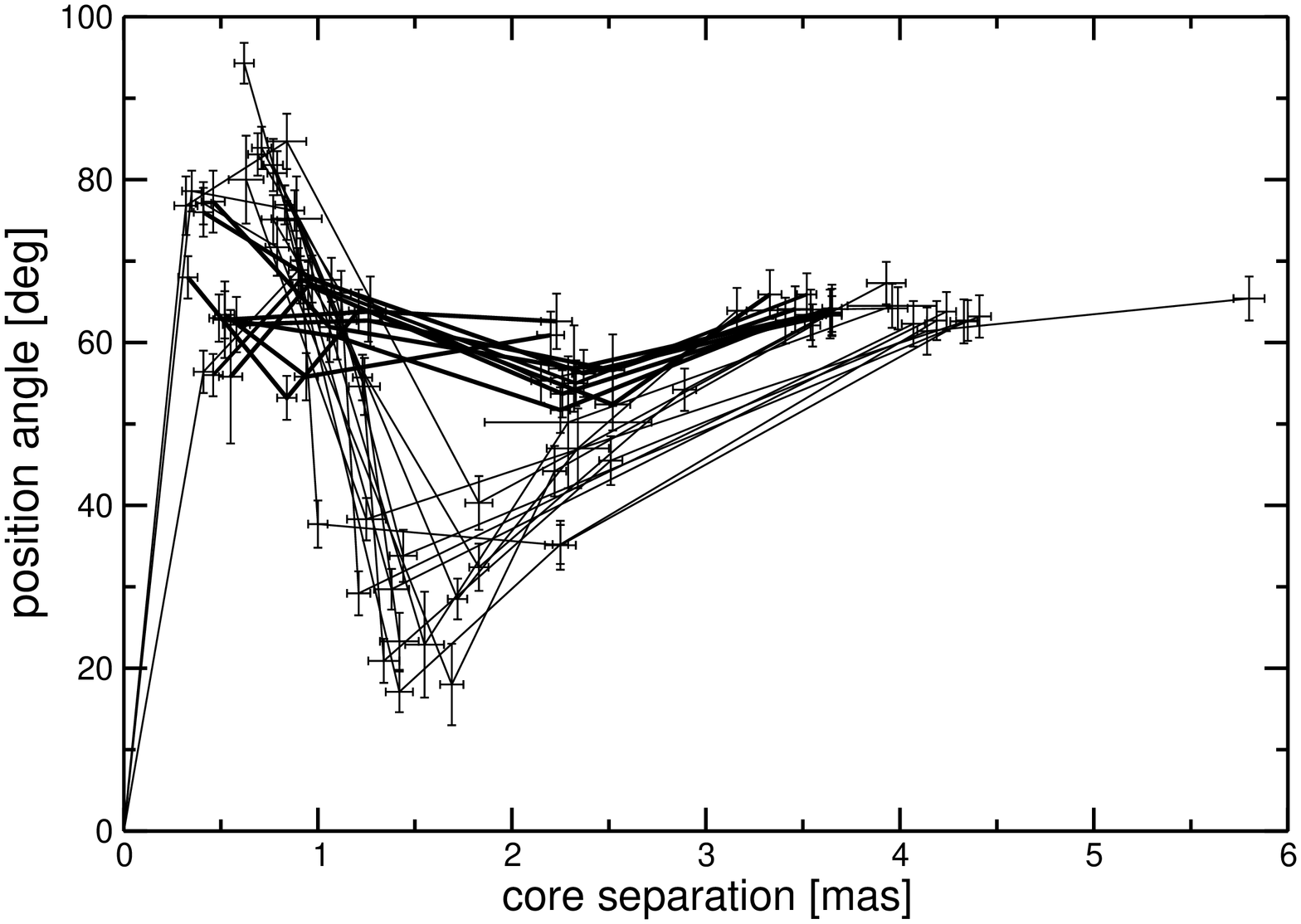}}
   \subfigure[]{\includegraphics[clip,angle=-90,width=8.5cm]{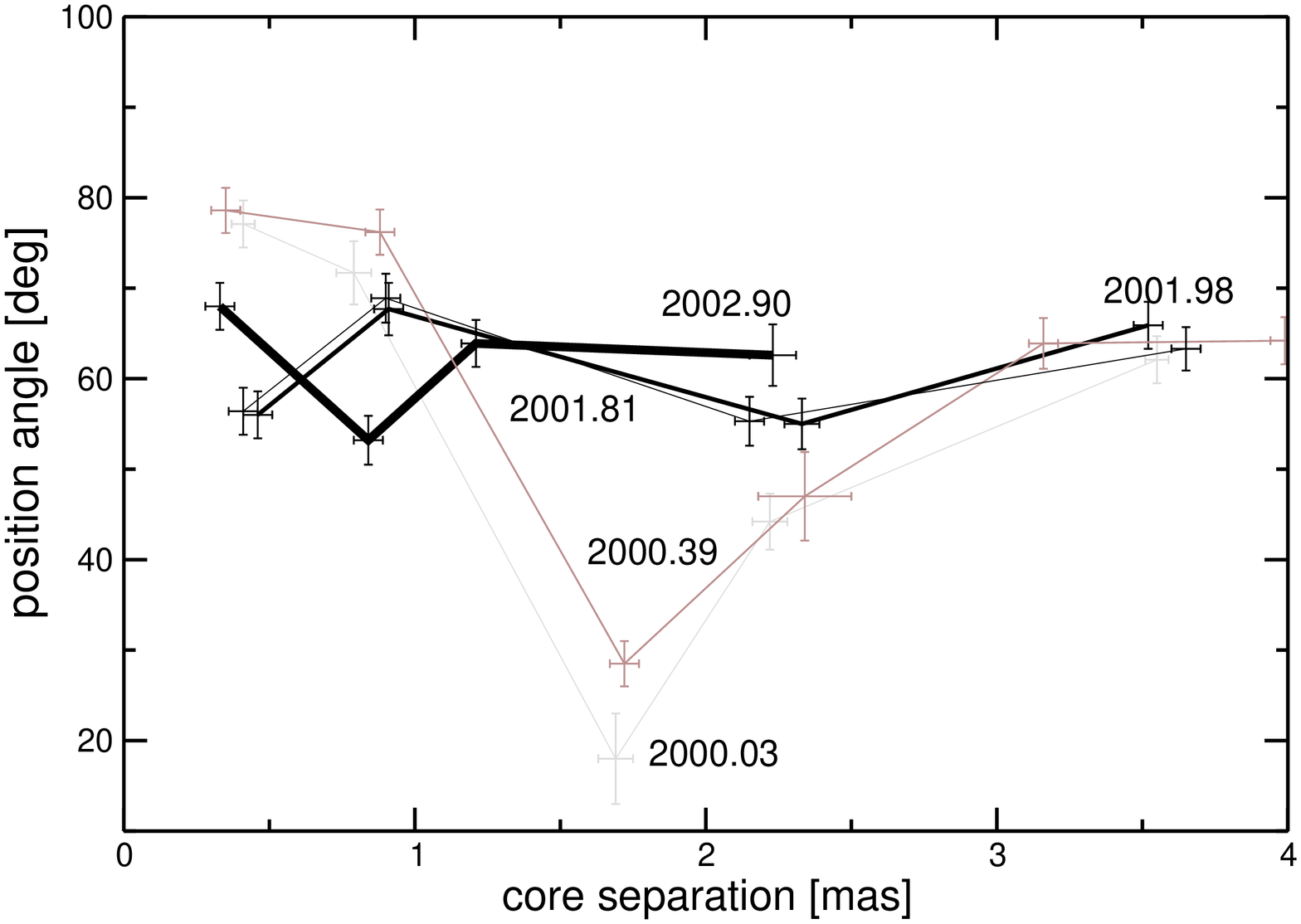}}\\
   \subfigure[]{\includegraphics[clip,angle=-90,width=8.5cm]{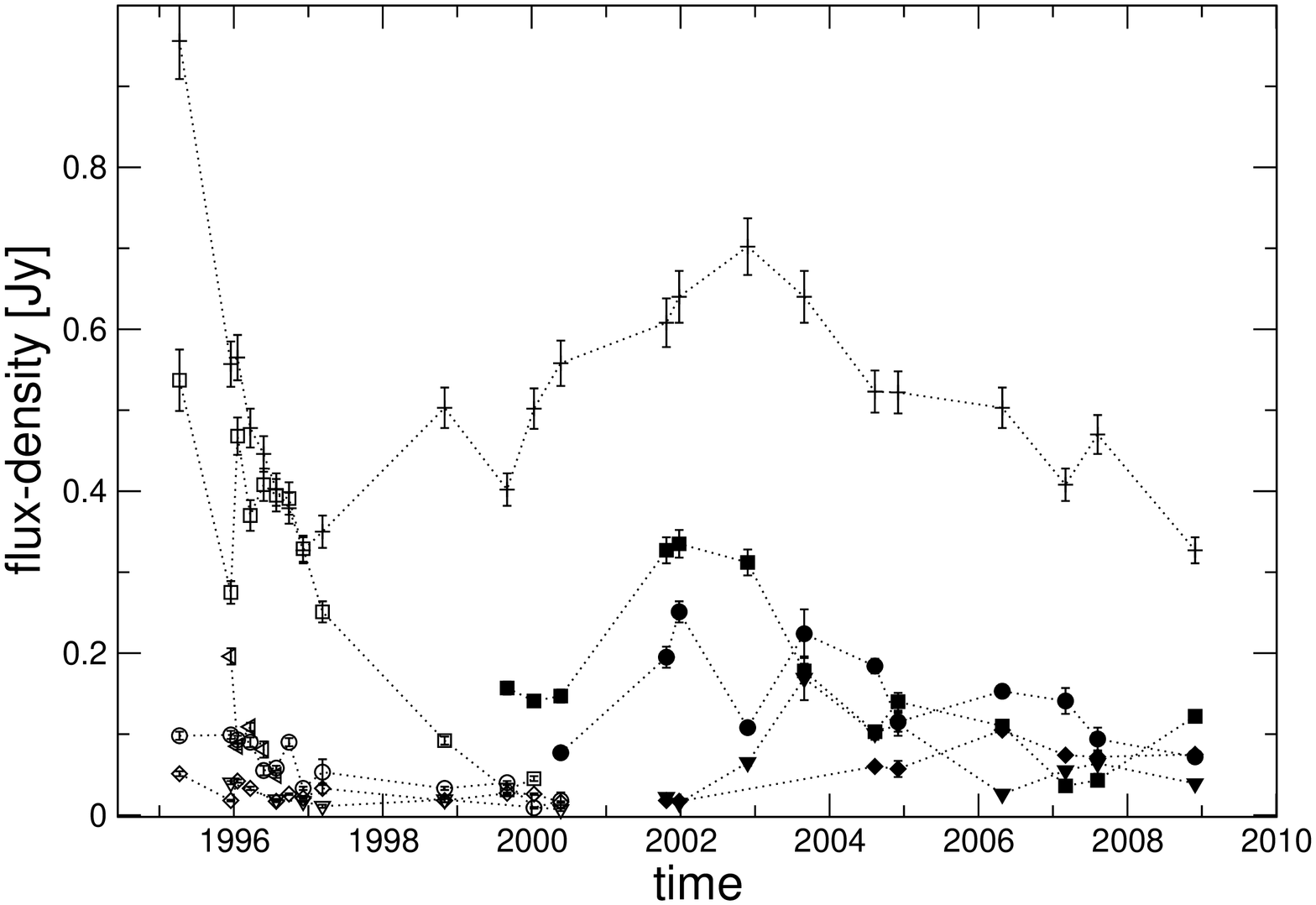}}
   \subfigure[]{\includegraphics[clip,angle=-90,width=8.5cm]{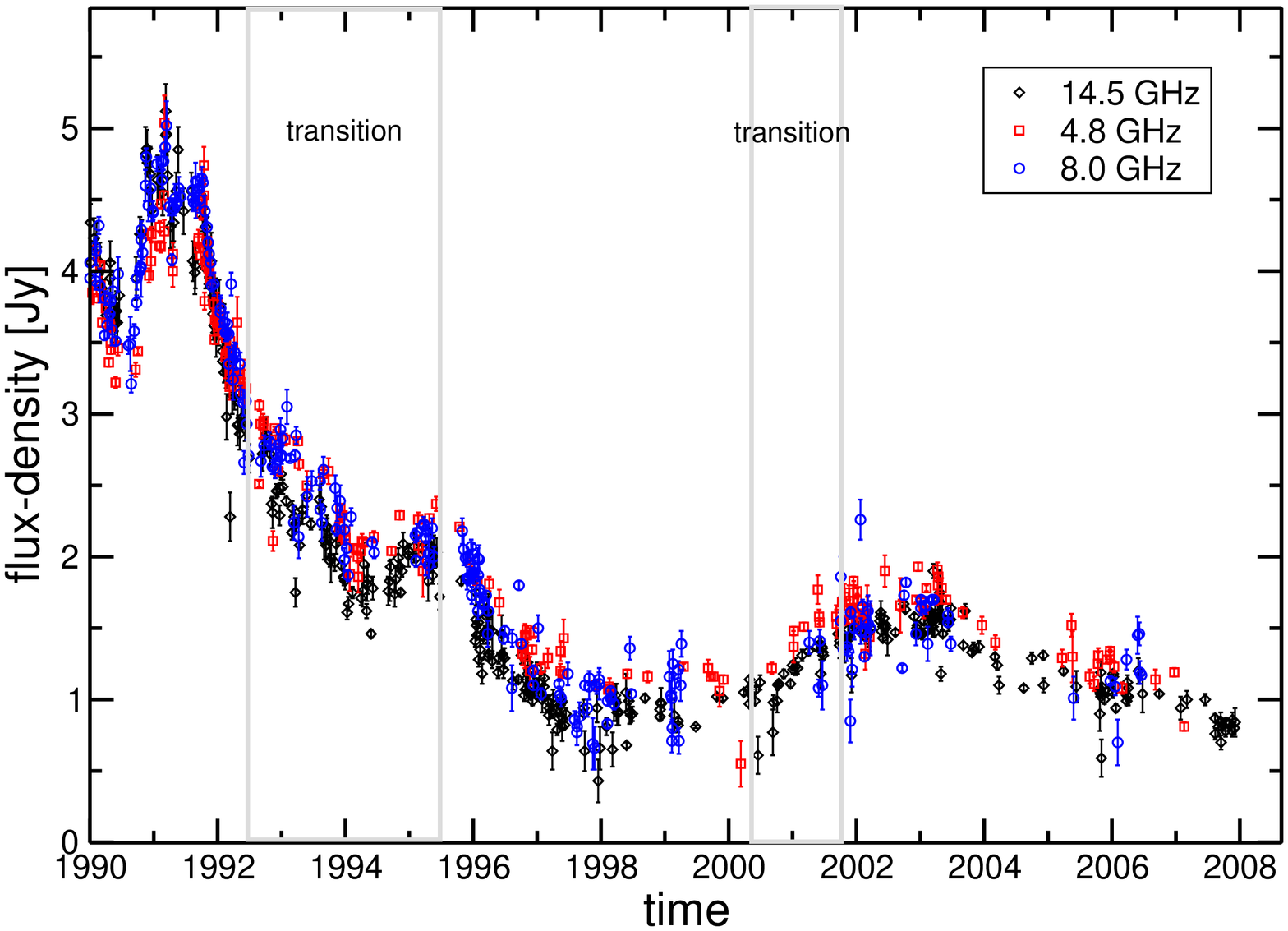}}\\
                \caption{a) Core separation of each VLBI component as 
function of time for both time 
intervals described in the text. Grey symbols represent weak components that could not be identified through the epochs. b) Position angle of each component as function of time, for both time intervals. c) The jet ridge-line for all the epochs
observed. d) Three representative jet ridge-lines selected from the two time intervals, for comparison. 
e) Flux-density evolution of the individual components and the core (upper 
data points). f) Radio flux-density light-curves from the Michigan monitoring 
program (UMRAO). Indicated are the two transition phases where the mode changes
(i.e., morphological transitions) occurred
(Sect. 3). }
                          \label{comparison}
                                    \end{figure*}
                            \clearpage
                            \pagebreak
			    \begin{figure}[htb]
			     \centering
			      \includegraphics[clip,angle=-90,width=9.5cm]{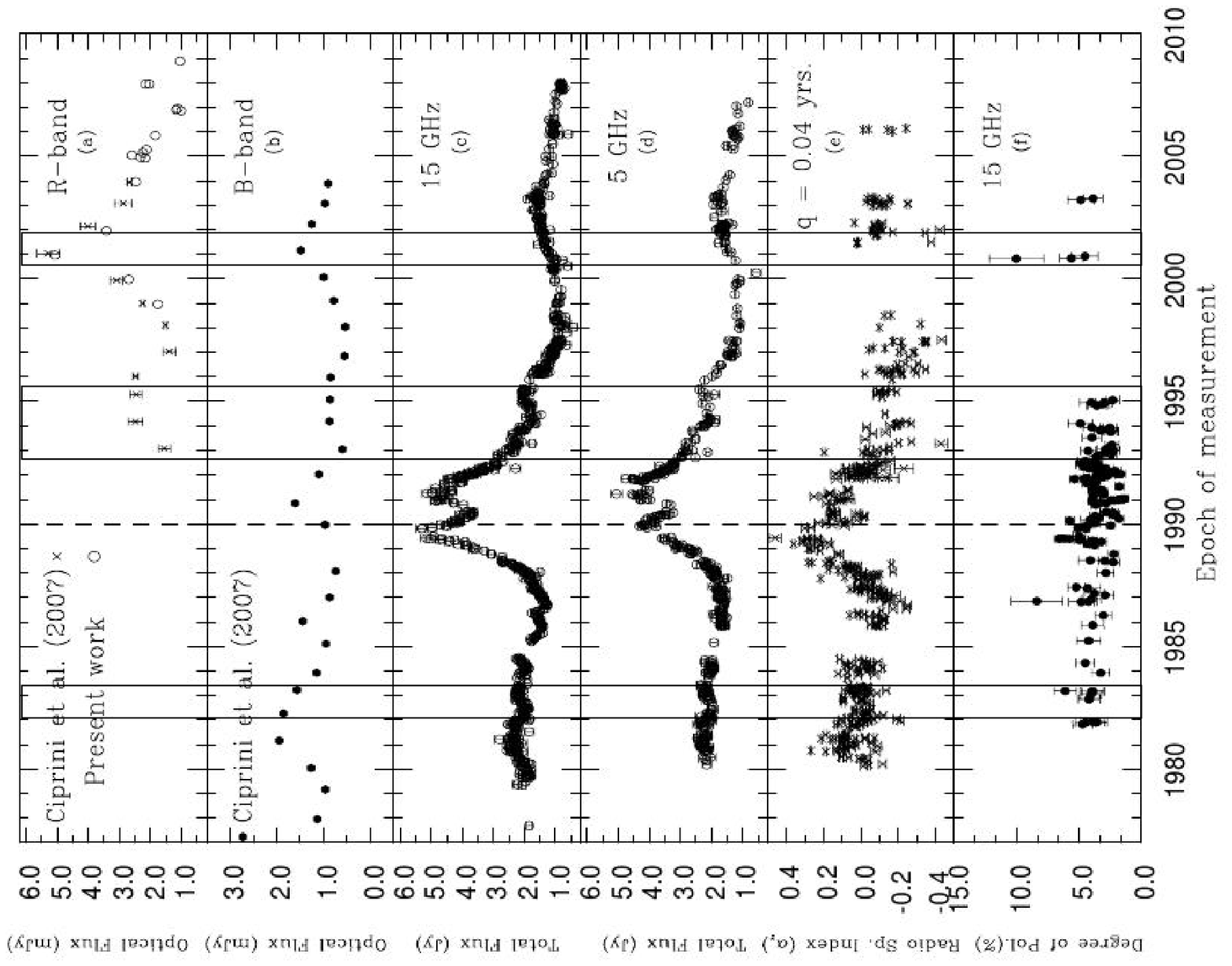}
			      \caption{The diagram, reproduced from Goyal et 
al. (2009), shows the multiple light-curves of PKS 0735+178.
Also displayed are the R-band light curve (a), 
the B-Band light curve (b), the UMRAO light curves at 15 and 5 GHz (c) \& (d) 
and the run of the radio spectral index (e). The plot (f) shows the variation 
of the percentage (linear) polarization at 15 GHz. The two pairs of vertical
lines bracket the structural transition phases discussed in the text.}
			      \label{optical}
			      \end{figure}

\subsection{Estimation of the uncertainties}
Although the use of VLBA data generally ensures a high quality, considerable 
variation can occur between the epochs in both the amount of data and its
quality. The uncertainties were determined individually for each epoch by 
comparing the model-fit parameters for the sets with different numbers of 
components ($\pm$1). For the uncertainties of the flux-density we assumed 
a minimum error of 5\%. The positional uncertainty was determined by 
averaging 1/5th of the beam size and the value determined by comparing several 
model-fit results. For the size we adopted an uncertainty of 10\%. The model 
fit parameters and their uncertainties are given in Tables \ref{modelfits1}
and \ref{modelfits2}. 
We also list there the epoch of observation, the flux-density of each fitted
component and its separation from the core, the corresponding position
angle, and the component size.

\section{Results}
\label{results}

\begin{figure*}[htb]
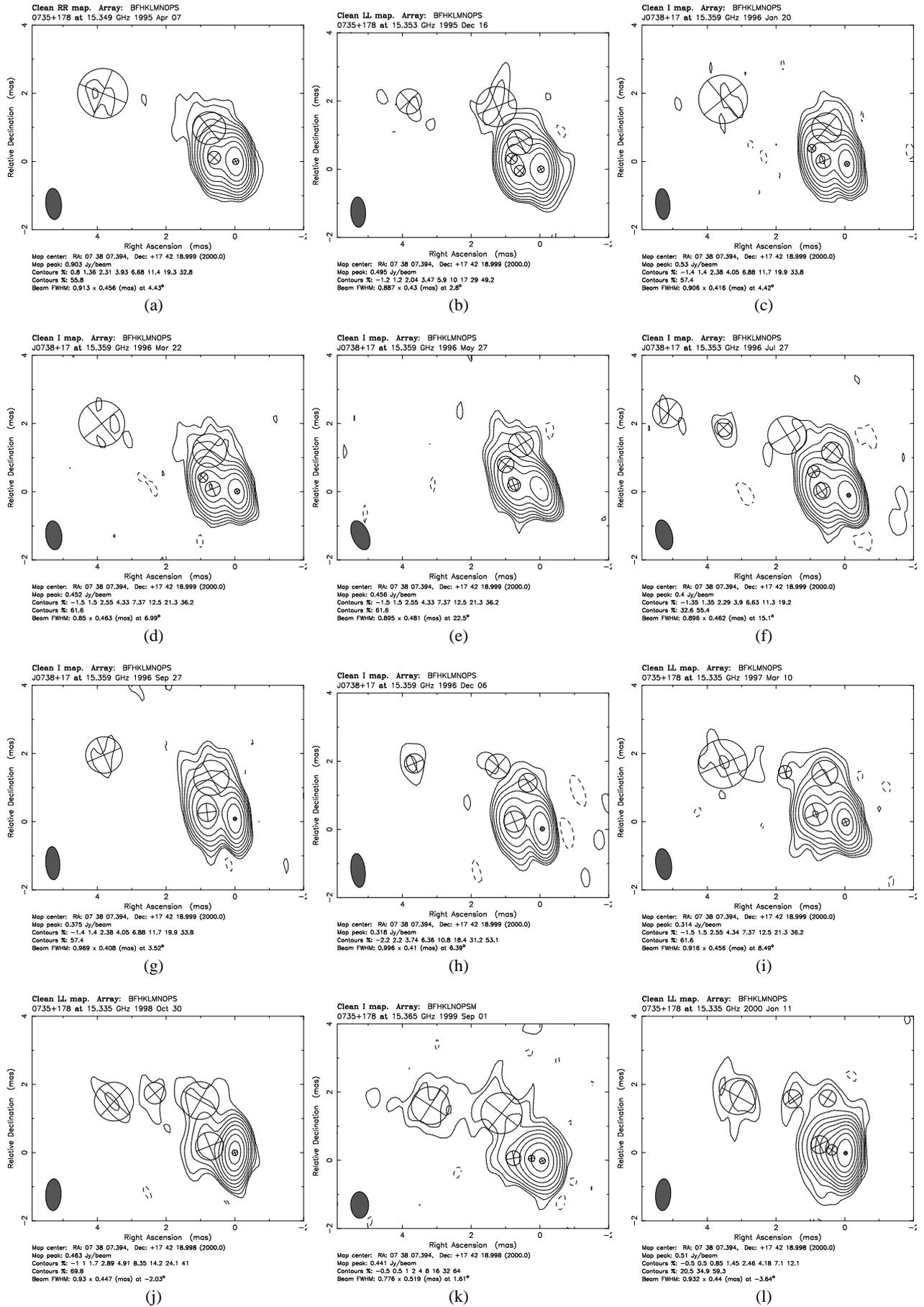

 \centering
     \subfigure[]{\includegraphics[clip,width=5.5cm]{13685f6a.ps}}
        \subfigure[]{\includegraphics[clip,width=5.5cm]{13685f6b.ps}}
        \subfigure[]{\includegraphics[clip,width=5.5cm]{13685f6c.ps}}\\
             \subfigure[]{\includegraphics[clip,width=5.5cm]{13685f6d.ps}}
               \subfigure[]{\includegraphics[clip,width=5.5cm]{13685f6e.ps}}
                      \subfigure[]{\includegraphics[clip,width=5.5cm]{13685f6f.ps}}\\
                          \subfigure[]{\includegraphics[clip,width=5.5cm]{13685f6g.ps}}
                               \subfigure[]{\includegraphics[clip,width=5.5cm]{13685f6h.ps}}
                 \subfigure[]{\includegraphics[clip,width=5.5cm]{13685f6i.ps}}\\
                \subfigure[]{\includegraphics[clip,width=5.5cm]{13685f6j.ps}}
                \subfigure[]{\includegraphics[clip,width=5.5cm]{13685f6k.ps}}
               \subfigure[]{\includegraphics[clip,width=5.5cm]{13685f6l.ps}}
            \caption{The VLBI images superimposed with model-fitting results are shown for all epochs between 1995.27 and 2000.03.}
              \label{images1}
         \end{figure*}

\begin{figure*}[htb]
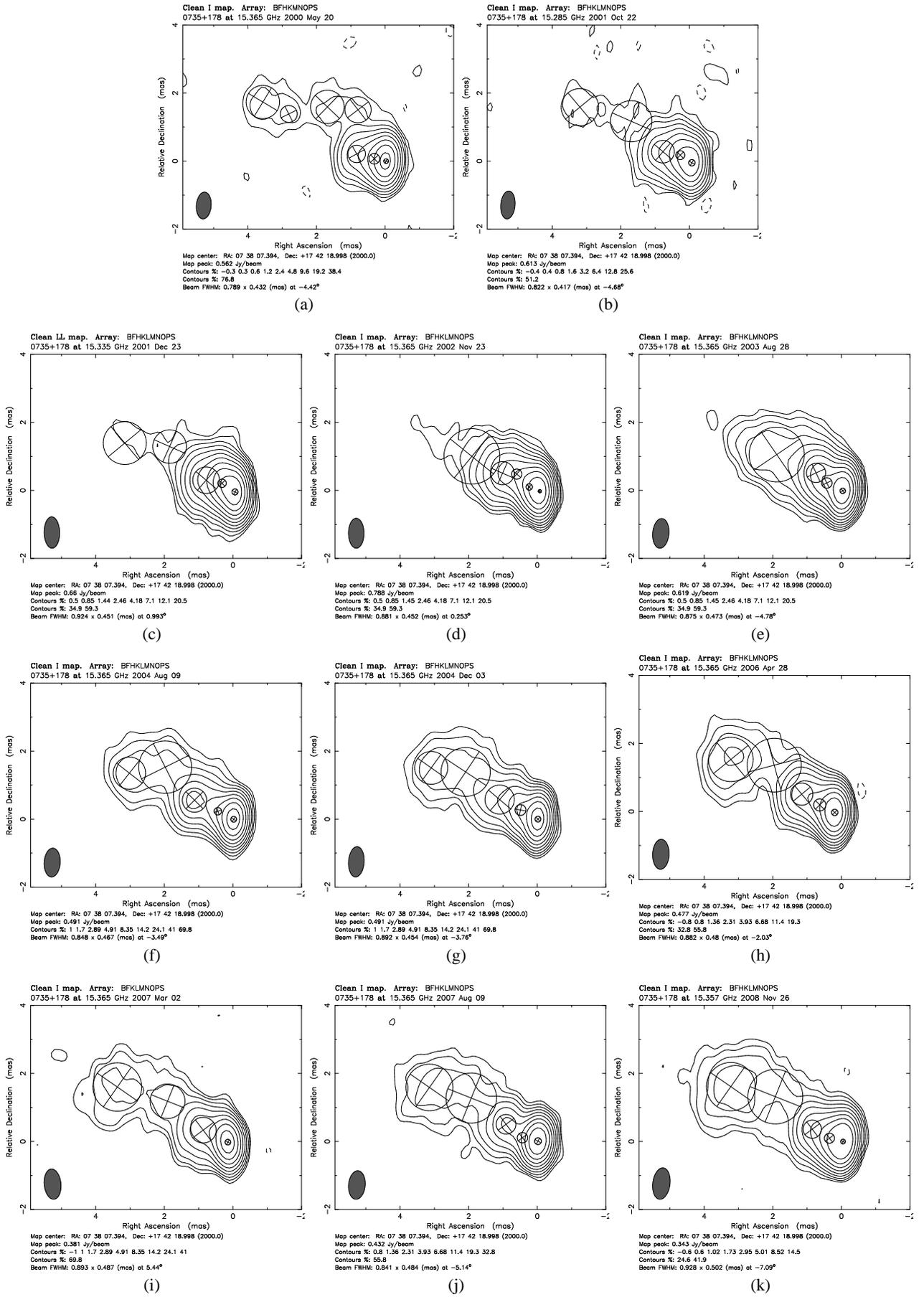

  \centering
                \subfigure[]{\includegraphics[clip,width=5.5cm]{13685f7a.ps}}
                \subfigure[]{\includegraphics[clip,width=5.5cm]{13685f7b.ps}}\\
     \subfigure[]{\includegraphics[clip,width=5.5cm]{13685f7c.ps}}
       \subfigure[]{\includegraphics[clip,width=5.5cm]{13685f7d.ps}}
  \subfigure[]{\includegraphics[clip,width=5.5cm]{13685f7e.ps}}\\
       \subfigure[]{\includegraphics[clip,width=5.5cm]{13685f7f.ps}}
    \subfigure[]{\includegraphics[clip,width=5.5cm]{13685f7g.ps}}
           \subfigure[]{\includegraphics[clip,width=5.5cm]{13685f7h.ps}}\\
           \subfigure[]{\includegraphics[clip,width=5.5cm]{13685f7i.ps}}
           \subfigure[]{\includegraphics[clip,width=5.5cm]{13685f7j.ps}}
               \subfigure[]{\includegraphics[clip,width=5.5cm]{13685f7k.ps}}
    \caption{The VLBI images superimposed with model-fitting results are shown for all epochs between 2001.98 and 2008.91.}
            \label{images2}
               \end{figure*}

Figures~\ref{images1} and \ref{images2} show the images derived by us,
with the model-fit components superimposed. It is evident that the jet's
morphology undergoes a drastic change during the time span covered by
our observations. In Fig.~\ref{images1} the jet shows the twice bent 
structure - the {\it staircase} mode, whereas in Fig.~\ref{images2} the 
jet structure is seen to have become straight.
Figure~\ref{comparison}~a) shows the variation
of core separations between the epochs 1995.27 and 2008.91 for the different components.
Striking variations in the jet kinematics are seen during
the period 1995 -- 2000.4. Four jet components clearly show outward directed 
motion with apparent superluminal speeds, while one component shows an 
inward motion (directed towards the core) also with an apparent superluminal 
speed. The apparent speeds derived from linear regressions to the 
component positions are given in Table \ref{speeds}. From 2001.8 on, only 
four components can be identified in the VLBI jet,
and the jet components are found to be stationary (Figs.~\ref{images1} and 
\ref{images2}).\\
Figure~\ref{comparison}~b) presents the  position angle 
(relative to the core) at different epochs for each component. Again we find a significant 
change to have occured between 2000.4 and 2001.8. Between 1995 and 2000 the 
position angle distribution is quite broad with values ranging from 
$\sim$10$^{\circ}$--100$^{\circ}$, which amounts to a wide spread of
over $\sim$90$^{\circ}$. After 2000.4, the position angle distribution is seen to become much narrower, with a total spread of 
just about 20$^{\circ}$ around a mean value of $\sim$65$^{\circ}$.

In Fig.~\ref{comparison}~c) the relation between the core separation and 
position angle is plotted for each of the 23 epochs. This plot more clearly depicts the two contrasting trends present in the VLBI images. Thus, a 
different pattern is observed for the epochs before 2000.4 and after 
2000.4. Prior to 2000.4 all the jet ridge lines show a similar form, with a
clear minimum at a core separation of $\sim$1.5mas, where the position angle 
drops to $\sim 20^{\circ}$. After 2000.4, the form of the jet ridge lines
becomes almost linear. \\
Thus, both the jet morphology and kinematics are found to change remarkably
between 1995 and 2009, as summarized below:\\

\begin{enumerate}
\item In the {\it staircase} jet mode five radio knots/components move 
with apparent superluminal speeds either away from the core (four components), 
or towards it (one component). The position angle distribution is broad 
($\sim90^{\circ}$). The jet ridge lines for all epochs before 2000.4 are 
similar, with a clear minimum in position angle. \\
\item In the {\it straight} jet mode, separations of all four VLBI components
from the core remain essentially unchanged; their apparent speeds are 
sub-luminal. The position angle distribution is narrow ($\sim 20^{\circ}$). \\
  \end{enumerate}
In Fig.~\ref{comparison}~d) we present the same information as in 
Fig.~\ref{comparison}~c), but for selected epochs only: the two epochs before 
the mode change and the first three epochs following the mode change (2001.81,
 2001.98 and 2002.90). Here the different jet ridge lines for the two modes 
are seen more clearly. Figure~\ref{comparison}~e) displays the evolution of 
flux-densities of the individual components and the core; again significant 
changes are observed.
 \begin{table}
        \caption[]{Proper motion and apparent speeds of the individual jet 
components in 0735+178 (Sect. 3).}
           \label{speeds}
 \begin{tabular}{lccc}
 \hline\hline
              mode&Component& $\mu$ [mas/year]&$\beta_{\rm app}$ [$c$]      \\
              \noalign{\smallskip}
              \hline
              \noalign{\smallskip}
              {\it staircase}&c1 &0.03 $\pm$ 0.01&0.77 $\pm$ 0.26     \\
              &c2 & 0.46 $\pm$0.16& 11.87 $\pm$ 4.16   \\
              &c3& 0.22 $\pm$ 0.02 & 5.68 $\pm$ 0.52 \\
              &c4 &0.22 $\pm$ 0.04 & 5.68 $\pm$ 1.03  \\
              &c5 &-0.13 $\pm$ 0.03 & -3.36 $\pm$ 0.77   \\
  \hline
              {\it straight}&ca & 0.01 $\pm$ 0.01&  0.26$\pm$ 0.26  \\
              &cb & 0.02 $\pm$ 0.02&   0.52 $\pm$ 0.52  \\
              &cc & 0.01 $\pm$ 0.02&  0.26 $\pm$ 0.52  \\
              &cd &  0.01 $\pm$ 0.02&  0.26 $\pm$ 0.52   \\
              \noalign{\smallskip}
              \hline
\end{tabular}
     \end{table}

\subsection{Morphological transitions and flux-density variability}
The BL Lac 0735+178 is known for its pronounced flux variability in both 
optical and radio bands (Ciprini et al. 2007; Goyal et al. 2009). The origin 
of this variability, however, is still debated. To check for a possible 
correlation between the optical/radio flux-density variability and the 
morphological transitions witnessed in the VLBI observations, we show 
the light curves (Figs.~\ref{comparison}~f, \ref{optical}) with the epochs of the observed mode transitions 
indicated by pairs of grey and black lines, respectively.

\subsubsection{Radio flux variability}
In Fig.~\ref{comparison}~f) the intergrated flux-densities at 4.8, 8.0, and 
14.5 GHz are plotted, monitored under the University of Michigan programme (UMRAO). 
The grey stripes denote the two most recent mode transition events.
Based on the VLBI images reported here and in the literature (see above), 
one can deduce that the first of these mode transitions ({\it straight} to
{\it staircase}) occured sometime 
between mid-1992 and mid-1995, and hence its time-scale must be less than 
three years. Likewise, the latest mode transition ({\it staircase} to 
{\it straight}) we report in this paper took less than two years. The time interval between the two transitions is 
$\sim$7.5 years. 

In Fig.~\ref{optical} we show a plot adopted from Goyal et al. (2009) and 
indicate in it all three known mode transitions within pairs of vertical
black lines. The 2001 transition occured shortly before a clear radio flare. The 1994 transition occured shortly 
before a clear modest radio flare. The 1983 mode transition cannot be 
related to a significant radio flare. While the corelation between the times of the transititons and 
the optical flaring is convincing, the case is not so clear with regard to a possible correlation with radio flares. It seems that
for the last two transitions the radio flares are delayed with regard to the
optical flare, which could be explained by propagation effects within the source. Unfortunately only one VLBI
observation - the observation by Gabuzda et al. (1994) from 1990.47 - is available 
for the time of the most prominent radio flaring around 1990. This would help to understand the relation
between the radio flaring and the morphological changes. A simultaneous monitoring of the radio and optical
flux in combination with multi radio frequency VLBI observation is required for a better understanding of the emission processes.
\\

\subsubsection{Correlation with optical flux variability}
In Fig.~\ref{optical}(a, b) we reproduce the optical light curves from Goyal 
et al. (2009), on which we have marked the three known mode transitions
mentioned above. Both {\it staircase} to {\it straight} mode changes 
clearly take place near the brightness maxima. The optical flares coincident
with these two transitions 
are more prominent compared to the middle flare coincident with the 
opposite mode transition. We thus find that
a strong correlation exists between the structural change in
the parsec-scale jet and the optical light curve. To our knowledge, such a 
correlation has not been pointed out earlier for any jet, and this 
underscores the importance of long-term, coordinated VLBI and optical 
monitoring of blazars.\\  
The origin of this conspicuous correlation in this source needs to be 
investigated in greater detail. This correlation could be explained 
by a small change in the direction to the line of sight. This would change the
amplification of intrisinc jet bending and also allow for a more direct view into the
inner regions of the AGN. This ``geometric'' explanation should be explored 
further.
While more subtle interpretations are possible, a simple possibility 
to understand this correlated radio-optical behavior is that the 
optical flare manifests an impulse by which a large amount of 
kinetic power (momentum flux) is injected by the central engine into the 
nuclear jet, as a result of which the jet gets stretched into a linear shape.
This explanation is related to the question of higher stability of powerful jets and the
collimation process of jets in general (e.g., Nakamura \& Meier 2009). 
The observed time lag between the optical flare and the disappearance
of the jet's kink could in fact be quite small ($<$ 1 yr), since the jet
flow in blazars is directed close to the line-of-sight, which is also the 
direction of 
propagation of the optical signal recorded. An alternative scenario is 
presented in Sect. \ref{helix}. Either of these two possible models will also be 
required to explain the disappearance of superluminal motion following the mode
transition to a straight jet.\\
The proposed energetic impulse that straightens the kinky jet 
could originate from the core. Assuming that 0735+178 is a helical jet,
whose components are therefore seen at different viewing angles at the times 
of different mode transitions, then this could explain the connection to the 
radio flares as well. 
The radio flare would then not be caused by the ejection of a new component - 
which is usually assumed - but by the Doppler factor changing with the
viewing angle.

\section{Discussion}

\subsection{Morphological mode transitions}

PKS 0735+178 showed a kinky jet trajectory in the VLBI observations in 
December 1981 and a straight jet trajectory in June 1983. Therefter, 
sometime between mid-1992 and mid-1995, the morphology changed back to  
kinky with two sharp bends of $\sim 90^{\circ}$ ({\it staircase} mode),
within 2 msec of the core. 
Between 2000.4 and 2001.8 the jet trajectory changed once again to the
{\it straight} mode, as found in the present work. Thus, three clear
mode changes of the pc-scale trajectory of this jet were observed. 
The time gaps between these transitions are about 11.3 years and 7.2 years. 
A simple scenario for the mode transition from kinky to straight jet
was outlined in the previous section, taking a clue from the
observed coincidences of the morphological transition events 
(kinky to straight jet) with the optical flares of this blazar. Below 
we present a kinematic model for the latest mode transition revealed 
by our analysis of the VLBI observations taken at 23 epochs between
1995 to 2009 (Sect. \ref{helix}). \\ 

\subsection{Jet kinematics}
\label{kinematics}
During the {\it staircase} mode, outward and inward apparent superluminal 
motion is observed. During the {\it straight} mode, the separations
of various knots/components from the core remained essentially unchanged.
Whether these components are the {\bf same} or different, cannot be 
decided with absolute certainty. However, should the mode transition be
caused by a changing viewing angle - a possibility argued below - the
components could be the same.\\
A similar phenomenon - rather stationary components relative to the core,
but moving perpendicular to the jet ridge line - was studied by us 
in the two prominent BL Lac objects, 1803+784 (Britzen et al., 2009) and 
0716+714 (Britzen et al., 2009). Over a time of $\sim$ 20 years we do find 
rather slow apparent motion or stationarity for nearly all of the components 
in the pc-scale jet of 1803+784. However, the components are seen to move
significantly in position angle measured from the core. The entire ridge
line of the jet thus evolves in the transverse direction, changing from a
straight line to an almost sinusoidal form and back again to a straight 
line. These significant jet ridge-line changes occur 
on a timescale of $\sim$ 8 years. A similar phenomenon was studied 
by us in the highly variable BL Lac 0716+714. Here again we find rather 
stationary components that do, however, show motion in position angle
and, once again, the jet's ridge line evolves. We expect that the stationary components can be explained in terms of a geometric origin. In the past, various mechanisms were proposed to explain slow or subluminal apparent motions. Smaller apparent speeds can be obtained by a flux-weighted integration of the apparent motions
 over the entire jet width (Gopal-Krishna et al. 2004). For a brief review we refer to Britzen et al. (2008) and references therein.\\ Different studies have reached opposite conclusions on whether BL Lac objects typically have slow or fast jet components (e.g., Gabuzda et al. 1994; Wehrle et al. 1992; Ghisellini et al. 1993; Jorstad et al. 2001). The CJF (Caltech-Jodrell Bank Flat-Spectrum) survey of 293 AGN (Britzen et al. 2008) shows some tendency for smaller apparent velocities of BL Lac objects compared to quasars. However, the small size of the BL Lac sample handicaps the comparison. 
\\
\subsection{A ``hybrid'' between quasars and BL Lac objects?}
As noted in the previous subsection, there has been an ongoing debate in
recent years whether BL Lac objects and Quasars reveal similar kinematic properties or whether the apparent speeds differ significantly. According to the general picture of the AGN paradigm, fast apparent speeds with values up to $\sim$30{\it c} can be observed (e.g., Piner et al. 2006) and in almost all cases the motion is directed away from the assumed stationary core. This is in contrast to what we find in a growing number of BL Lac objects. Based on our analysis of 0735+178 presented here and the detailed studies of 1803+784 and 0716+714, it is possible that the kinematic properties are significanlty different. 
0735+178 is a ``hybrid'' with regard to the mode changes and the accompanying kinematics - it shows `quasar'-like fast apparent motions at certain periods of time and then again `slow' motion (as we find in BL Lac objects) at other times. Thus this source switches between the different states that we tend to associate with 
quasars or BL Lac objects. Due to this atypical behavior, 0735+178 can help us to understand the difference between these two types of AGN. It is often assumed that the basic difference between these two classes is based on a different line of sight to the object, with the BL Lacs being seen under the smallest angle to the line of sight (Unified Schemes). Within this scheme, it is then quite conceivable that the mode changes are caused by small changes in the angle to the line of sight, e.g., the jet bends slightly towards or away from us. Undoubtedly, 0735+178 is an extreme object since these changes of the angle can not be very considerable in the source frame. Thus, 0735+178 might already be oriented in a critical angle towards us. This would explain why this is the only source so far where these mode fluctutations have been observed.  
\begin{figure}
       \centering
       \includegraphics[clip,height=6cm]{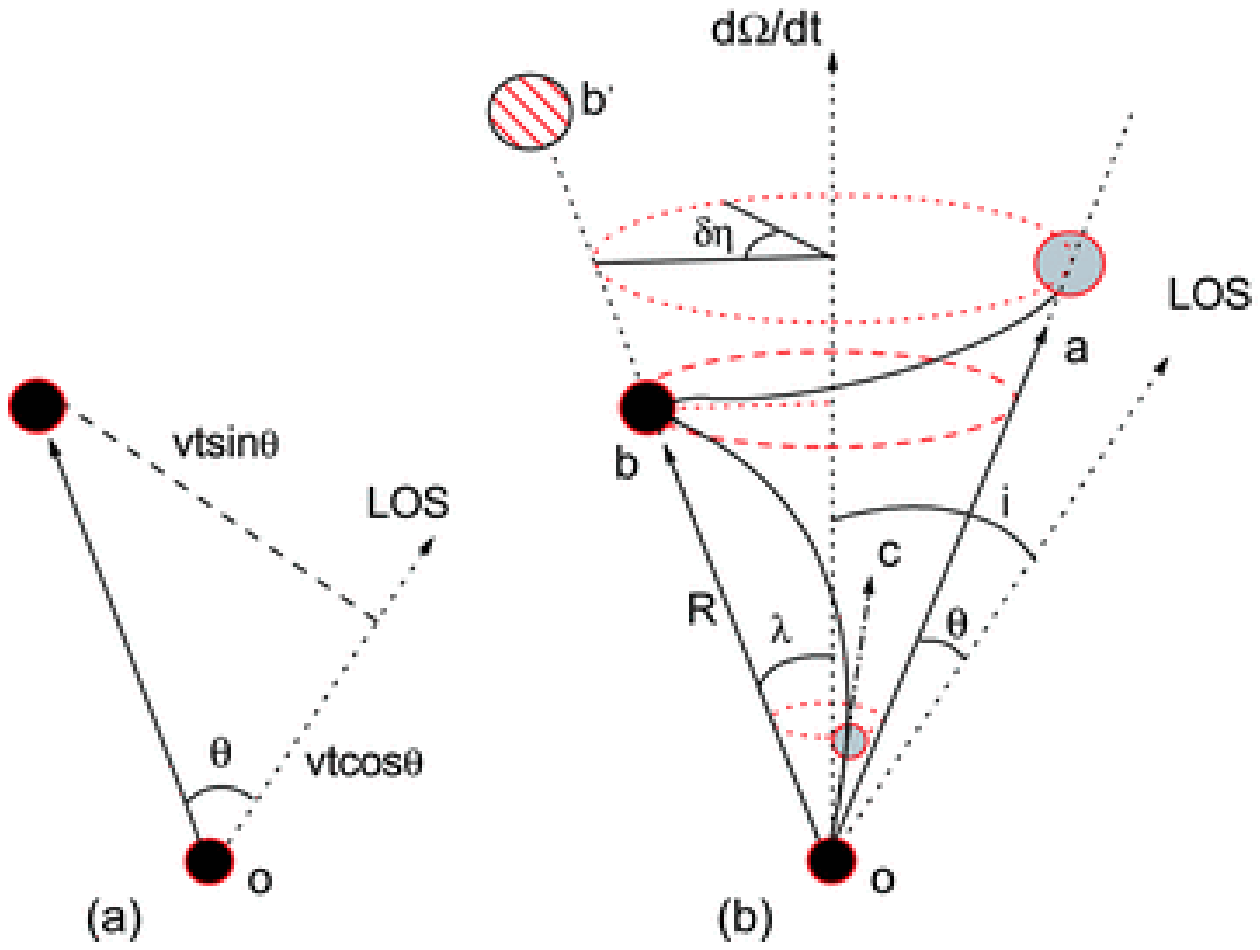}
       \caption{\label{Fig:gong}  This figure is reproduced from Gong (2008) and shows
       the difference between ballistic and non-ballistic motion.
       The model of ballistic motion plus precession is represented by the solid spiral 
       connecting the features {\it a}, {\it b}, and {\it c}. Denoted by the dashed
       ellipse through feature b is the non-ballistic model which precesses with a constant distance
       to the core.}
       \end{figure}

\subsection{Helix or pressure gradients in the external medium?}
\label{helix}
Gong (2008) proposed a non-ballistic model (see Fig.~\ref{Fig:gong}), in which a continuous jet 
produces discrete hotspots via the interaction of the relativistic outflow 
with the ambient medium. Thereby significant bulk energy of the jet is 
transferred to the swept up material. The precession of the jet axis 
results in the motion of hotspots at different directions along the 
precession cone, which resembles a hotspot moving along a ring-like 
trajectory. Significant bulk energy loss may occur repeatedly for a 
continuous jet with a fixed axis, leading to several distinct hotspots 
at different distances from the core. The precession of such a continuous 
jet results, on the other hand, in the motion of hotspot at different 
rings and with different precession phases (a detailed analysis is in 
preparation).

Under this scenario, the rectangular structure is just a state of the 
motion of these hotspots in different rings, which can appear and disappear 
depending on the precession phases of these hotspots and their projections 
on the sky plane. Thus, conceivably, the non-ballistic model of Gong (2008) 
can provide a fairly simple explanation for the appearance of the {\it staircase} 
mode.

\begin{figure}
       \centering
\includegraphics[clip,height=8cm]{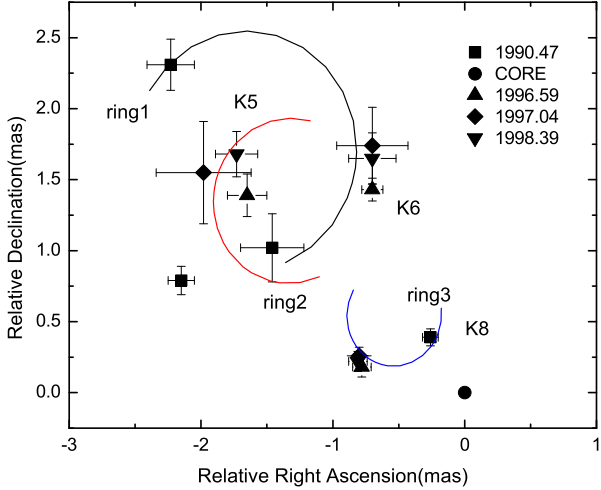}
\caption{\label{Fig:f1} 
The observations of the BL Lac object 0735+178 at 8.4 GHz from G$\acute{\rm o}$mez
et al.(2001) as well as the curves fitted to the data. }
\end{figure}

Figure~\ref{Fig:f1} shows the observation of the BL Lac object 0735+178 
at 8.4 GHZ by G$\acute{\rm o}$mez et al. (2001) as well as the curves 
fitted to the data. The inner components K8, K6 and K5 - as defined in G$\acute{\rm o}$mez 
et al. (2001) - are fitted within the non-ballistic model (Gong 2008) by 
three solid curves (corresponding to ring~1, ring~2 and ring~3, respectively) 
at different distances to the core. The evolution of the three curves
is clockwise, all of them start in 1988.47 and end in 2002.97.

Under the simplest treatment, the non-ballistic model just gives the same 
opening angle for the precession cone for the three rings, 
$\lambda_1=\lambda_2=\lambda_3$. However, in this case the morphology 
of PKS 0735+178 cannot be fitted well.

Thanks to the scenario of the pressure gradients in the external medium 
through which the jet propagates (G$\acute{\rm o}$mez et al. 2001), 
it is reasonable to assume that $\lambda_1$,  $\lambda_2$, $\lambda_3$
can deviate from each other by a few degrees. Under these circumstances,
the evolution of the {\it staircase} mode can be well fitted, as shown 
in Fig.~\ref{Fig:f1}. The input parameters to this model are given in 
Table \ref{nonballistic}.

\begin{table}
\caption{The parameters applied to fit the data of 0735+178 in the 
non-ballistic model.}             
\label{nonballistic}      
\centering                          
\begin{tabular}{cccccc}        
\hline\hline                 
$\dot{\Omega}$ & i & $\xi$ & $\lambda_1$ & $\lambda_2$ &  $\lambda_3$   \\ 
15.0 & 9.8 & 225.6 & 3.5 & 3.0 &  4.6  \\
 \hline
$\eta_1$ & $\eta_2$ &  $\eta_3$ & $R_1$ & $R_2$ &  $R_3$  \\
165.2 & 21.7 & -53.0 & 14.0 & 11.1 &  4.5 \\
\hline

 \hline
       \end{tabular}
             \end{table}

The precession speed, $\dot{\Omega}$, is in units of deg/yr, the
distance of a hotspot to the core, $R_i$, is in units of mas, and
all other parameters are in units of degree. $i$ is the inclination angle
between the jet's rotation axis and the line of sight; $\xi$ is the
coordinate transformation angle; $\lambda_i$ is the opening angle of the 
precession cone; $\eta_i$ is the initial phase.

\begin{figure*}
       \centering
\includegraphics[clip,height=14cm]{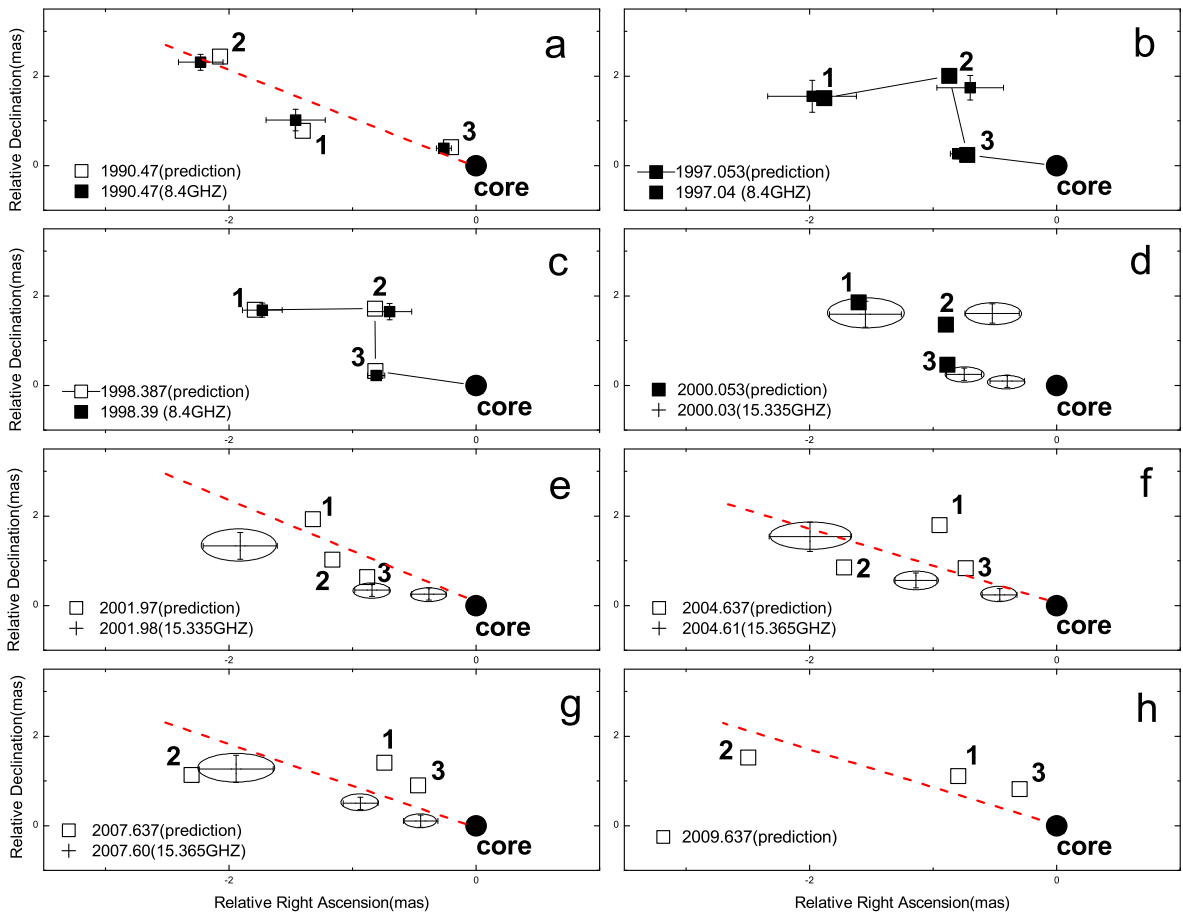}
\caption{\label{Fig:f2} The observed morphology of 0735+178 is
indicated by filled squares and ellipses, corresponding to the VLBI
observations at 8.4 GHz (1990-1998) and at 15 GHz (2000-2007).  The
evolution of the trajectory of hotspots (hollowed square) from
1990 to 2009 is reproduced in terms of the non-ballistic model.}
\end{figure*}

Although Fig.~\ref{Fig:f2} shows the fit to the 8.4 GHz observations
of G$\acute{\rm o}$mez et al. (2001), the theoretical trajectory
(Fig.~\ref{Fig:f2} h) can be extended to up to 2009.63. By doing this,
the disappearance of the {\it staircase} observed at 15.335 GHz and
15.365 GHz (symbolized by the ellipses) at the end of 2001 can also
be explained as shown in Fig.~\ref{Fig:f2}. This figure shows the evolution of
the jet ridge line based on the observed morphology (filled filled squares and ellipses) and
the model (open squares). For the time of 2009.637 the model prediction is shown in the last panel of this figure.
This figure distinctly shows the different morphological
phases of the jet ridge line as they are observed and how well they can be reproduced
within the model.

Consequently, the {\it staircase} mode occurs when the hotspot precesses 
to certain phases so that the projected appearance is rectangular. 
Likewise, the {\it staircase} mode disappears when the hotspot precesses
to other phases where the projection deviates from the rectangular image.
Obviously, the precession of a hotspot varies the angle of the jet axis 
from the line of sight. This leads to a varying Doppler boosting, which 
explains the observed radio flux-density changes.
The non-ballistic model, as outlined before, can explain the observed phenomena.
When interpreted in the context of the ballistic model, the two sharp
apparent bends of 90 deg should move outward as a whole relative to the
core. Since each component in the {\it staircase} mode corresponds to a
knot ejected previously, it should not have remained stationary.
The {\it staircase} mode may induce pressure gradients in the medium
through which the jet propagates, triggered by its own interaction with
the ambient medium. In this case, gradual changes in the position and
curvature of the jet bends near K8 and K6 with time can be expected
(G$\acute{\rm o}$mez et al. 2001). Plausibly, such a scenario could be
responsible for the small angle variations of the jet. However, to
explain the two 90 degree bends would require very special conditions
in the ambient medium, making it hard to understand why such an ambient
medium existed only during 1992-2000, but was absent before mid-1992 and
again after 2001.
As pointed out by G$\acute{\rm o}$mez et al. (2001), the moving-component
scenario suggests very non-ballistic trajectories for K5 and K6. But there
are still unanswered questions concerning the origin of these non-ballistic
trajectories. Are they caused by continuous or discret ejection events,
and what role, if any, is played by the jet precession? Further detailed investigations and modeling
is required to answer these questions.

\subsection{Timescales: Flux-density variability and mode change}
The three drastic mode changes are clearly correlated 
with maxima in the optical light-curve. Thus we assume that the optical 
emission (most likely emergent from the core) causes the change in the 
trajectory. In two cases, a major long-term radio flux outburst occurs
shortly after a mode change has appeared and this can be explained within 
the non-ballistic motion scenario of hotspots outlined in the previous 
subsection. In the optical emission, Fan et al. (1997) and Qian \& Tao (2004) 
found periodic features of 13.8-14.2 years. This is roughly twice the 
time interval we find here between the two mode transitions. A physical 
connection between the morphological changes (mode changes) and the flux 
variability in the radio and the optical regime is thus likely.\\ 

According to conventional wisdom, radio outbursts are explained as correlated with the ejection of a new radio component from the core.
However, in PKS 0735+178 we found a similar number of components over a 
time span of roughly 14 years and no new components were unequivocally detected during this period. In addition, we find a correlation 
with the morphological 
changes described as mode changes. Furthermore, we are able to explain
the flux-density changes within the model by Gong et al. (2008) of 
non-ballistically moving hotspots. Thus, it can be inferred  
that in PKS 0735+178
ejection of VLBI components from the core is {\it not} required to explain 
the observed (quasi-periodic) radio flaring. \\ 

We found similar results for the other well-known blazars, 1803+784 
and 0716+714 - the long-term radio variability can be correlated with 
morphological changes in the nuclear jet and is hence also likely 
to be of geometric origin.  We thus 
find in three BL Lac objects supporting evidence for a correlation between 
long-term radio variability and morphological changes that are likely to be 
of geometric origin. Additional contribution to morphological changes,
particularly the more abrupt ones, could
come from short bursts of power injected into the jet by the central engine,
manifested by the optical flaring events.

\section{Conclusions}

\begin{itemize}
\item{We report a third (and the most clear) event of structural mode 
transition in the nuclear jet of the BL Lac object PKS 0735+178. In
particular, we find that the twice sharply bent trajectory of the 
nuclear jet transformed into a straight jet sometime between 200.4 and
2001.8.} 

\item{From a literature search we further found that a similar 
transition from a kinky to a straight jet is present in the VLBI 
observations of this source performed at 5 GHz in December 1981 and June 1983.} 

\item{Contrasting kinematic properties are found associated with the two
morphological modes of the nuclear jet:
in the {\it staircase} mode apparent superluminal motion of the knots in
the jets is observed, either directed away from the core or towards it.
In the {\it straight} mode, however, the knots remain fairly stationary 
relative to
the core. Their position angle distribution is broad in the former case 
($\sim$ 90$^{\circ}$) and narrow in the latter ($\sim$ 20$^{\circ}$).}

\item{The time interval between the mode changes is $\sim$ 11.3 years 
between the first two mode changes and $\sim$ 7.2 years between the last
two mode changes.}

\item{Both mode changes, from kinky to straight jet, are found to coincide 
with individual, well-defined peaks in the optical light-curve. This points
to a causal connection between the optical flaring and the structural 
changes in the nuclear jet.}

\item{The last two mode changes are seen to occur shortly before major 
radio outbursts.}

\item{Remarkably, no new radio knot/component was observed during 
the entire time span (1995.27 to 2008.91) covered by the VLBI 
images presented in this study - the components in the {\it staircase}-mode 
are fairly unambiguously identifiable with the set of components 
witnessesd in the {\it straight} mode. A smooth transition was 
shown in this paper, which could be explained by a (most likely curling) 
jet with a changing viewing angle.}

\item{The kinematics of the VLBI knots were modeled within the 
non-ballistic motion scenario (Gong et al. 2008).}

\item{The long-term radio flux variability can be explained by Doppler 
boosting effects arising from the precession of the nuclear jet.}

\item{The model of non-ballistic motion of hotspots explains the variability 
of flux-density and morphology, the significant morphological changes and the 
unusual kinematics of the nuclear jet in PKS0735+178. }

\item{Similar kinematic properties - fairly stationary components - and a 
correlation between periodic ridge line changes and the long-term 
radio outbursts were found by some of us for two other well-known 
blazars: 1803+784 and 0716+714.}
\end{itemize}
\begin{acknowledgements}
This research made use of data from the MOJAVE database that is maintained by the MOJAVE team. This research made use of data from the University of Michigan Radio Astronomy Observatory, which is supported by the National Science Foundation and by funds from the University of Michigan. This research made use of the NASA/IPAC Extragalactic Database (NED) which is operated by the Jet Propulsion Laboratory, California Institute of Technology, under contract with the National Aeronautics and Space Administration. The National Radio Astronomy Observatory is a facility of the National Science Foundation operated under cooperative agreement by Associated Universities, Inc.
The research of the authors Biping Gong and J.W.Zhang is supported in part by the National Natural Science Foundation of China, under grands 10778712.

\end{acknowledgements}

	       \clearpage
\tabcolsep0.6mm
\hspace*{-2.5cm}
\begin{table}
\caption{Model-fit results}             
\label{modelfits1}      
\centering                          
\begin{tabular}{ccccccc}        
\hline\hline                 
Epoch & Flux-density [Jy]& core sep. [mas] & pos. a. [deg]&Maj. a. [mas] & Id. \\    
\hline                        
1995 Apr 7& 0.956$\pm$ 0.047& 0.00$\pm$ 0.00&  0.0$\pm$  0.0&0.17$\pm$0.02&k \\
& 0.537$\pm$ 0.038& 0.63$\pm$ 0.09& 80.0$\pm$  5.4&0.37$\pm$0.04&c1\\
& 0.098$\pm$ 0.005& 1.25$\pm$ 0.10& 38.3$\pm$  2.6&0.96$\pm$0.10&c3\\
& 0.051$\pm$ 0.003& 4.35$\pm$ 0.06& 62.7$\pm$  2.5&1.47$\pm$0.15&c5\\
\\
1995 Dec 15& 0.557$\pm$ 0.028& 0.00$\pm$ 0.00&  0.0$\pm$  0.0&0.19$\pm$0.02&k \\
& 0.275$\pm$ 0.014& 0.62$\pm$ 0.05& 94.3$\pm$  2.5&0.32$\pm$0.03&c1\\
& 0.196$\pm$ 0.010& 0.91$\pm$ 0.05& 70.1$\pm$  2.7&0.33$\pm$0.03&c2\\
& 0.099$\pm$ 0.005& 1.00$\pm$ 0.05& 37.7$\pm$  2.9&0.73$\pm$0.07&c3\\
& 0.040$\pm$ 0.002& 2.25$\pm$ 0.08& 35.1$\pm$  3.0&1.18$\pm$0.12&c4\\
& 0.018$\pm$ 0.001& 4.33$\pm$ 0.07& 62.6$\pm$  2.7&0.74$\pm$0.07&c5\\
\\
1996 Jan 19 & 0.565$\pm$ 0.028& 0.00$\pm$ 0.00&  0.0$\pm$  0.0&0.15$\pm$0.02&k \\
& 0.468$\pm$ 0.023& 0.69$\pm$ 0.05& 83.1$\pm$  2.6&0.43$\pm$0.04&c1\\
& 0.085$\pm$ 0.004& 1.12$\pm$ 0.05& 66.1$\pm$  2.7&0.23$\pm$0.02&c2\\
& 0.093$\pm$ 0.005& 1.21$\pm$ 0.06& 29.2$\pm$  2.7&0.87$\pm$0.09&c3\\
& 0.042$\pm$ 0.002& 4.07$\pm$ 0.06& 62.3$\pm$  2.8&1.42$\pm$0.14&c5\\
\\
1996 Mar 22 & 0.478$\pm$ 0.024& 0.00$\pm$ 0.00&  0.0$\pm$  0.0&0.15$\pm$0.02&k \\
& 0.370$\pm$ 0.019& 0.71$\pm$ 0.05& 83.9$\pm$  2.6&0.44$\pm$0.04&c1\\
& 0.109$\pm$ 0.005& 1.07$\pm$ 0.05& 67.7$\pm$  2.7&0.28$\pm$0.03&c2\\
& 0.090$\pm$ 0.006& 1.44$\pm$ 0.07& 33.8$\pm$  3.2&0.99$\pm$0.10&c3\\
& 0.033$\pm$ 0.002& 4.41$\pm$ 0.06& 63.2$\pm$  2.6&1.34$\pm$0.13&c5\\
\\
1996 May 27 & 0.446$\pm$ 0.022& 0.00$\pm$ 0.00&  0.0$\pm$  0.0&0.00$\pm$0.10&k \\
& 0.408$\pm$ 0.020& 0.77$\pm$ 0.05& 81.8$\pm$  3.2&0.38$\pm$0.04&c1\\
& 0.081$\pm$ 0.010& 1.24$\pm$ 0.08& 54.6$\pm$  3.5&0.48$\pm$0.05&c2\\
& 0.055$\pm$ 0.006& 1.42$\pm$ 0.10& 23.3$\pm$  3.5&0.73$\pm$0.07&c3\\
\\
1996 Jul 27& 0.402$\pm$ 0.020& 0.00$\pm$ 0.00&  0.0$\pm$  0.0&0.11$\pm$0.10&k \\
& 0.395$\pm$ 0.020& 0.79$\pm$ 0.05& 80.8$\pm$  2.7&0.48$\pm$0.05&c1\\
& 0.049$\pm$ 0.003& 1.23$\pm$ 0.05& 55.7$\pm$  2.8&0.35$\pm$0.04&c2\\
& 0.058$\pm$ 0.003& 1.34$\pm$ 0.08& 20.9$\pm$  2.7&0.62$\pm$0.06&c3\\
& 0.019$\pm$ 0.001& 2.51$\pm$ 0.06& 45.5$\pm$  3.0&1.12$\pm$0.11&c4\\
& 0.017$\pm$ 0.001& 4.14$\pm$ 0.05& 61.4$\pm$  2.9&0.52$\pm$0.05&c5\\
& 0.012$\pm$ 0.001& 5.80$\pm$ 0.08& 65.4$\pm$  2.7&0.86$\pm$0.09&c6\\
\\
1996 Sep 27& 0.379$\pm$ 0.019& 0.00$\pm$ 0.00&  0.0$\pm$  0.0&0.10$\pm$0.10&k \\
& 0.391$\pm$ 0.020& 0.83$\pm$ 0.05& 76.9$\pm$  2.4&0.53$\pm$0.05&c1\\
& 0.090$\pm$ 0.005& 1.38$\pm$ 0.09& 29.7$\pm$  2.5&1.03$\pm$0.10&c3\\
& 0.026$\pm$ 0.001& 4.24$\pm$ 0.05& 63.8$\pm$  2.4&1.07$\pm$0.11&c5\\
\\
1996 Dec 06& 0.327$\pm$ 0.016& 0.00$\pm$ 0.00&  0.0$\pm$  0.0&0.12$\pm$0.10&k \\
& 0.329$\pm$ 0.016& 0.84$\pm$ 0.05& 75.0$\pm$  2.4&0.62$\pm$0.06&c1\\
& 0.033$\pm$ 0.002& 1.42$\pm$ 0.07& 17.1$\pm$  2.5&0.55$\pm$0.06&c3\\
& 0.017$\pm$ 0.001& 2.25$\pm$ 0.04& 35.2$\pm$  2.4&0.73$\pm$0.07&c4\\
& 0.021$\pm$ 0.001& 4.19$\pm$ 0.05& 62.7$\pm$  2.4&0.57$\pm$0.06&c5\\
\\
1997 Mar 10 & 0.350$\pm$ 0.020& 0.00$\pm$ 0.00&  0.0$\pm$  0.0&0.22$\pm$0.02&k \\
& 0.251$\pm$ 0.013& 0.89$\pm$ 0.13& 75.2$\pm$  5.2&0.66$\pm$0.07&c1\\
& 0.053$\pm$ 0.016& 1.55$\pm$ 0.10& 22.9$\pm$  6.5&0.76$\pm$0.08&c3\\
& 0.011$\pm$ 0.002& 2.29$\pm$ 0.43& 50.2$\pm$  8.1&0.39$\pm$0.04&c4\\
& 0.033$\pm$ 0.006& 3.96$\pm$ 0.23& 64.5$\pm$  2.7&1.43$\pm$0.14&c5\\
\\
1998 Oct 30& 0.503$\pm$ 0.025& 0.00$\pm$ 0.00&  0.0$\pm$  0.0&0.17$\pm$0.02&k \\
& 0.092$\pm$ 0.005& 0.77$\pm$ 0.06& 75.1$\pm$  3.0&0.82$\pm$0.08&c1\\
& 0.033$\pm$ 0.002& 1.83$\pm$ 0.05& 32.4$\pm$  2.9&1.03$\pm$0.10&c3\\
& 0.019$\pm$ 0.003& 2.89$\pm$ 0.06& 54.2$\pm$  2.6&0.90$\pm$0.09&c4\\
& 0.018$\pm$ 0.002& 3.93$\pm$ 0.10& 67.3$\pm$  2.6&1.01$\pm$0.10&c5\\
\\
1999 Sep 1 & 0.402$\pm$ 0.020& 0.00$\pm$ 0.00&  0.0$\pm$  0.0&0.17$\pm$0.02&k \\
& 0.157$\pm$ 0.008& 0.32$\pm$ 0.06& 76.8$\pm$  3.6&0.19$\pm$0.02&ca\\
& 0.031$\pm$ 0.006& 0.84$\pm$ 0.10& 84.7$\pm$  3.4&0.43$\pm$0.04&c1\\
& 0.040$\pm$ 0.002& 1.83$\pm$ 0.07& 40.3$\pm$  3.3&1.20$\pm$0.12&c3\\
& 0.027$\pm$ 0.001& 3.64$\pm$ 0.06& 63.5$\pm$  3.0&1.10$\pm$0.11&c5\\
\\
2000 Jan 11& 0.502$\pm$ 0.025& 0.00$\pm$ 0.00&  0.0$\pm$  0.0&0.10$\pm$0.10&k \\
& 0.141$\pm$ 0.007& 0.41$\pm$ 0.04& 77.1$\pm$  2.6&0.31$\pm$0.03&ca\\
& 0.045$\pm$ 0.003& 0.79$\pm$ 0.06& 71.7$\pm$  3.5&0.52$\pm$0.05&c1\\
& 0.009$\pm$ 0.002& 1.69$\pm$ 0.06& 18.0$\pm$  5.0&0.50$\pm$0.05&x \\
& 0.009$\pm$ 0.001& 2.22$\pm$ 0.06& 44.2$\pm$  3.1&0.54$\pm$0.05&c3\\
& 0.026$\pm$ 0.001& 3.55$\pm$ 0.04& 62.1$\pm$  2.6&1.06$\pm$0.11&c5\\
 \hline                                   
       \end{tabular}
             \end{table}
\begin{table}
\caption{Model-fit results continued}             
\label{modelfits2}      
\centering                          
\begin{tabular}{ccccccc}        
\hline\hline                 
Epoch & Flux-density [Jy]& core sep. [mas] & pos. a. [deg]&Maj. a. [mas]&Id. \\    
\hline                        
2000 May 20& 0.558$\pm$ 0.028& 0.00$\pm$ 0.00&  0.0$\pm$  0.0&0.14$\pm$0.01&k \\
& 0.147$\pm$ 0.008& 0.35$\pm$ 0.05& 78.6$\pm$  2.5&0.31$\pm$0.03&ca\\
& 0.077$\pm$ 0.004& 0.88$\pm$ 0.05& 76.2$\pm$  2.5&0.51$\pm$0.05&cb\\
& 0.009$\pm$ 0.001& 1.72$\pm$ 0.05& 28.5$\pm$  2.5&0.77$\pm$0.08&x \\
& 0.018$\pm$ 0.010& 2.34$\pm$ 0.16& 47.0$\pm$  4.9&0.98$\pm$0.10&c3\\
& 0.007$\pm$ 0.002& 3.16$\pm$ 0.05& 63.9$\pm$  2.8&0.50$\pm$0.05&c4\\
& 0.016$\pm$ 0.001& 3.99$\pm$ 0.05& 64.2$\pm$  2.6&0.98$\pm$0.10&c5\\
\\
2001 Oct 22 & 0.608$\pm$ 0.030& 0.00$\pm$ 0.00&  0.0$\pm$  0.0&0.19$\pm$0.02&k \\
& 0.327$\pm$ 0.016& 0.41$\pm$ 0.05& 56.4$\pm$  2.6&0.26$\pm$0.03&ca\\
& 0.195$\pm$ 0.013& 0.90$\pm$ 0.05& 68.9$\pm$  2.7&0.66$\pm$0.07&cb\\
& 0.022$\pm$ 0.001& 2.15$\pm$ 0.05& 55.3$\pm$  2.7&1.21$\pm$0.12&cc\\
& 0.018$\pm$ 0.001& 3.65$\pm$ 0.05& 63.3$\pm$  2.4&1.09$\pm$0.11&cd\\
\\
2001 Dec 23& 0.640$\pm$ 0.032& 0.00$\pm$ 0.00&  0.0$\pm$  0.0&0.18$\pm$0.10&k \\
& 0.335$\pm$ 0.017& 0.46$\pm$ 0.05& 56.0$\pm$  2.6&0.25$\pm$0.03&ca\\
& 0.251$\pm$ 0.013& 0.91$\pm$ 0.05& 67.7$\pm$  2.9&0.78$\pm$0.08&cb\\
& 0.015$\pm$ 0.001& 2.33$\pm$ 0.06& 55.0$\pm$  2.8&0.98$\pm$0.10&cc\\
& 0.017$\pm$ 0.001& 3.52$\pm$ 0.05& 65.9$\pm$  2.6&1.26$\pm$0.13&cd\\
\\
2002 Nov 23& 0.702$\pm$ 0.035& 0.00$\pm$ 0.00&  0.0$\pm$  0.0&0.09$\pm$0.10&k \\
& 0.312$\pm$ 0.016& 0.33$\pm$ 0.05& 68.0$\pm$  2.6&0.19$\pm$0.02&ca\\
& 0.108$\pm$ 0.005& 0.84$\pm$ 0.05& 53.2$\pm$  2.7&0.30$\pm$0.03&cb\\
& 0.214$\pm$ 0.011& 1.21$\pm$ 0.05& 63.9$\pm$  2.6&0.68$\pm$0.07&x \\
& 0.065$\pm$ 0.004& 2.23$\pm$ 0.08& 62.6$\pm$  3.4&1.63$\pm$0.16&cc\\
\\
2003 Aug 28& 0.640$\pm$ 0.032& 0.00$\pm$ 0.00&  0.0$\pm$  0.0&0.16$\pm$0.02&k \\
& 0.178$\pm$ 0.009& 0.52$\pm$ 0.05& 63.4$\pm$  2.9&0.31$\pm$0.03&ca\\
& 0.224$\pm$ 0.030& 0.94$\pm$ 0.06& 55.8$\pm$  2.9&0.56$\pm$0.06&cb\\
& 0.169$\pm$ 0.027& 2.20$\pm$ 0.07& 60.9$\pm$  2.9&1.61$\pm$0.16&cc\\
\\
2004 Aug 9& 0.523$\pm$ 0.026& 0.00$\pm$ 0.00&  0.0$\pm$  0.0&0.17$\pm$0.02&k \\
& 0.103$\pm$ 0.008& 0.52$\pm$ 0.06& 62.8$\pm$  4.7&0.22$\pm$0.02&ca\\
& 0.184$\pm$ 0.009& 1.27$\pm$ 0.08& 63.8$\pm$  4.3&0.71$\pm$0.07&cb\\
& 0.101$\pm$ 0.008& 2.52$\pm$ 0.09& 52.4$\pm$  3.2&1.54$\pm$0.15&cc\\
& 0.060$\pm$ 0.003& 3.33$\pm$ 0.06& 65.9$\pm$  3.0&0.95$\pm$0.10&cd\\
\\
2004 Dec 2& 0.522$\pm$ 0.026& 0.00$\pm$ 0.00&  0.0$\pm$  0.0&0.17$\pm$0.02&k \\
& 0.140$\pm$ 0.011& 0.58$\pm$ 0.07& 62.2$\pm$  3.4&0.32$\pm$0.03&ca\\
& 0.115$\pm$ 0.012& 1.26$\pm$ 0.06& 62.7$\pm$  2.6&0.85$\pm$0.09&cb\\
& 0.112$\pm$ 0.014& 2.52$\pm$ 0.06& 56.6$\pm$  4.4&1.44$\pm$0.14&cc\\
& 0.057$\pm$ 0.010& 3.46$\pm$ 0.09& 64.1$\pm$  2.8&0.97$\pm$0.10&cd\\
\\
2006 Apr 28& 0.503$\pm$ 0.025& 0.00$\pm$ 0.00&  0.0$\pm$  0.0&0.19$\pm$0.02&k \\
& 0.110$\pm$ 0.006& 0.49$\pm$ 0.05& 63.0$\pm$  2.9&0.36$\pm$0.04&ca\\
& 0.153$\pm$ 0.008& 1.10$\pm$ 0.05& 60.7$\pm$  2.8&0.65$\pm$0.07&cb\\
& 0.026$\pm$ 0.001& 2.25$\pm$ 0.05& 51.7$\pm$  2.8&1.58$\pm$0.16&cc\\
& 0.105$\pm$ 0.005& 3.41$\pm$ 0.05& 62.7$\pm$  2.8&1.30$\pm$0.13&cd\\
\\
2007 Mar 2  & 0.408$\pm$ 0.020& 0.00$\pm$ 0.00&  0.0$\pm$  0.0&0.17$\pm$0.02&k \\
& 0.036$\pm$ 0.005& 0.55$\pm$ 0.06& 55.8$\pm$  8.2&0.04$\pm$0.10&ca\\
& 0.141$\pm$ 0.016& 0.95$\pm$ 0.08& 67.0$\pm$  2.9&0.66$\pm$0.07&cb\\
& 0.055$\pm$ 0.003& 2.26$\pm$ 0.06& 53.7$\pm$  2.9&1.11$\pm$0.11&cc\\
& 0.074$\pm$ 0.004& 3.65$\pm$ 0.05& 64.2$\pm$  2.9&1.39$\pm$0.14&cd\\
\\
2007 Aug 9& 0.470$\pm$ 0.024& 0.00$\pm$ 0.00&  0.0$\pm$  0.0&0.20$\pm$0.02&k \\
& 0.043$\pm$ 0.003& 0.46$\pm$ 0.06& 77.3$\pm$  3.8&0.30$\pm$0.03&ca\\
& 0.094$\pm$ 0.014& 1.06$\pm$ 0.05& 61.9$\pm$  4.3&0.58$\pm$0.06&cb\\
& 0.064$\pm$ 0.013& 2.32$\pm$ 0.13& 56.8$\pm$  5.3&1.48$\pm$0.15&cc\\
& 0.072$\pm$ 0.006& 3.54$\pm$ 0.08& 63.4$\pm$  3.1&1.37$\pm$0.14&cd\\
\\
2008 Nov 26 & 0.327$\pm$ 0.016& 0.00$\pm$ 0.00&  0.0$\pm$  0.0&0.14$\pm$0.10&k \\
& 0.122$\pm$ 0.006& 0.41$\pm$ 0.05& 76.0$\pm$  3.0&0.29$\pm$0.03&ca\\
& 0.072$\pm$ 0.004& 0.97$\pm$ 0.05& 67.8$\pm$  2.9&0.53$\pm$0.05&cb\\
& 0.039$\pm$ 0.002& 2.37$\pm$ 0.06& 56.2$\pm$  2.9&1.60$\pm$0.16&cc\\
& 0.075$\pm$ 0.004& 3.65$\pm$ 0.05& 63.7$\pm$  2.9&1.42$\pm$0.14&cd\\
     \hline                                   
      \end{tabular}
      \end{table}
\end{document}